\documentclass[12pt]{article}

\usepackage{graphicx}

\newcommand{\be}{\begin{equation}}
\newcommand{\ee}{\end{equation}}
\newcommand{\bea}{\begin{eqnarray}}
\newcommand{\eea}{\end{eqnarray}}
\newcommand{\beqn}{\begin{eqnarray}}
\newcommand{\eeqn}{\end{eqnarray}}
\newcommand{\ba}{\begin{array}}
\newcommand{\ea}{\end{array}}

\newcommand{\noi}{\noindent}
\newcommand{\ra}{\rightarrow}
\newcommand{\epem}{e^+ e^-}
\newcommand{\epemt}{$e^+ e^- \;$}
\newcommand{\eetth}{$e^+ e^-\ra t \bar{t} H$}
\newcommand{\eettht}{$e^+ e^-\ra t \bar{t} H\;$}

\newcommand{\nnht}{$\epem \ra \nu \bar{\nu} H \;$}

\newcommand{\eezht}{$\epem \ra Z H \;$}

\def\sm{${\cal{S}} {\cal{M}}\;$}
\def\sms{${\cal{S}} {\cal{M}}$}

\newcommand{\beq}{\begin{equation}}
\newcommand{\eeq}{\end{equation}}

\begin{document}

\begin{titlepage}

\vspace*{0.1cm} \rightline{LAPTH-986}
\vspace*{0.1cm}\rightline{KEK-CP-141}
\vspace*{0.1cm}\rightline{KEK-TH-892}

\vspace{1mm}
\begin{center}

{\Large{\bf Full ${\cal O}(\alpha)$ electroweak and ${\cal
O}(\alpha_s)$ corrections to $e^+e^- \rightarrow t \bar{t} H $}}

\vspace{.5cm}

G. B\'elanger${}^{1)}$, F. Boudjema${}^{1)}$, J.
Fujimoto${}^{2)}$, T. Ishikawa${}^{2)}$, \\ T. Kaneko${}^{2)}$,
K. Kato${}^{3)}$, Y. Shimizu${}^{2)}$, Y. Yasui${}^{2)}$ \\

\vspace{4mm}

{\it 1) LAPTH${\;}^\dagger$, B.P.110, Annecy-le-Vieux F-74941,
France.}
\\ {\it
2) KEK, Oho 1-1, Tsukuba, Ibaraki 305--0801, Japan.} \\
{\it 3)
Kogakuin University, Nishi-Shinjuku 1-24, Shinjuku, Tokyo
163--8677, Japan.} \\

\vspace{10mm} \abstract{ We present the full ${{\cal O}}(\alpha)$
electroweak radiative corrections to  associated Higgs top pair
production in $e^+e^-$ collisions. We combine these results with a
new calculation of the full one-loop QCD corrections. The
computation is performed with the help of {\tt GRACE-loop}. We
find that the ${{\cal O}}(\alpha)$ correction can be larger than
the ${{\cal O}}(\alpha_s)$ corrections around the peak of the
cross section especially for a light Higgs mass. At threshold
these corrections are swamped by the QCD corrections which are
enhanced by the gluon Coulomb contribution. We have also
subtracted the complete QED corrections and expressed the genuine
weak correction both in the $\alpha$-scheme and the
$G_\mu$-scheme. This reveals that the genuine weak corrections are
not negligible and should be taken into account for a precision
measurement of this cross section and the extraction of the Yukawa
$t \bar t H$ coupling.}

\end{center}

\vspace*{\fill} $^\dagger${\small URA 14-36 du CNRS, associ\'ee
\`a l'Universit\'e de Savoie.} \normalsize
\end{titlepage}

\baselineskip 24pt

\section{Introduction}
After the discovery of the Higgs particle at the Large Hadron
Collider (LHC),  one of the most pressing issue is a proper
determination of the properties of this scalar since this would be
an important window on the mechanism of electroweak symmetry
breaking and the generation of mass. The LHC will be able to
furnish a few measurements on the couplings of the Higgs to
fermions and gauge bosons\cite{higgsproperties-lc-lhc} but the
most precise measurements will be performed in the clean
environment of a future $e^+e^-$ linear
collider(LC)\cite{NLC-report,tesla-report,GLC-report}. For
example, from the measurement of the Higgs decay branching ratios,
the Yukawa couplings of the light fermions can be determined at
the per-cent level at a $\sqrt{s}=300-500$ GeV linear
collider\cite{NLC-report,tesla-report,GLC-report} if the Higgs
boson has a mass below the $W$ pair threshold. This mass range for
the Higgs is consistent with the latest indirect precision
data\cite{mhlimit-03-2003} and covers the range predicted for the
lightest Higgs of the minimal supersymmetric model (MSSM). At a
TeV scale  LC, the associate production of a Higgs boson with a
top quark pair, \eetth, provides a direct information on the
top-Higgs Yukawa coupling. In the Standard Model, \sms, the cross
section of the $e^+e^- \rightarrow t\bar{t}H$ process reaches a
few fb, for a light Higgs and for centre of mass energies ranging
from 700 GeV to 1TeV. The expected accuracy for the determination
of the top-Higgs coupling is of order $5\%$ through the precision
measurement of this process at the LC
experiment\cite{NLC-report,tesla-report,GLC-report,
BattagliaDesch}. Considering such a high accuracy on the $t \bar t
H$  coupling one needs, on the theoretical side, to take into
account the effect of radiative corrections. The purpose of this
letter is to provide the full one-loop electroweak and QCD
corrections to \eettht for a standard model Higgs. Preliminary
results have already been presented in \cite{eetthloopfest2}.

The full tree-level calculation of \eettht has been done a decade
ago\cite{eetth-tree-full}. An earlier approximate calculation had
been performed by only taking into account the (dominant) photon
exchange diagrams\cite{eetth-tree-first}. QCD radiative
corrections have also been  performed by two groups. Dawson and
Reina investigated the ${\cal O}(\alpha_s)$ corrections but only
to the dominant photon exchange contribution\cite{eetthdawson}.
The full ${\cal O}(\alpha_s)$ correction has been computed by
Dittmaier {\it et al.}\cite{eetthzerwas}. Recently the
supersymmetric QCD corrections have also  been discussed in
\cite{eetthsusy}. On the other hand, due to the presence of the
large top Yukawa coupling, the electroweak radiative corrections
may  also be sizable. However, the calculation of the electroweak
radiative correction has been missing. We will present new results
of the full ${\cal O}(\alpha)$ corrections consisting of virtual
and soft corrections as well as hard photon radiation for the
process $e^+e^-\rightarrow t\bar{t}H$ in the \sm and will combine
this result with the QCD corrections.

\section{Grace-Loop and the calculation of \eettht}
 Our computation is performed with the help of {\tt GRACE-loop}.
This is a code for the automatic generation and calculation of the
full one-loop electroweak  radiative corrections in the \sms. It
has been successfully tested for a variety of one-loop $2\ra 2$
electroweak processes\cite{nlgfatpaper}. It also provided the
first results on the full one-loop radiative corrections to $e^+
e^- \ra \nu \bar{\nu} H$ \cite{eennhletter,eennhradcor2002} which
have recently been confirmed by an independent
calculation\cite{Dennereennh1}. For all electroweak processes we
adopt the on-shell renormalisation scheme according
to\cite{nlgfatpaper,eennhletter,kyotorc}. For each process some
stringent consistency checks are performed. The results, for the
part pertaining to the electroweak corrections,  are checked by
performing three kinds of tests at some random points in phase
space. For these tests to be passed one works in quadruple
precision. Details of how these tests are performed are given
in\cite{nlgfatpaper,eennhletter}. Here we only describe the main
features of these tests. We first check the ultraviolet finiteness
of the results. This test applies to the whole set  of the virtual
one-loop diagrams. In order to conduct this test we regularise any
infrared divergence by giving the photon a fictitious mass (we set
this at $\lambda=10^{-15}$GeV). In the intermediate step of the
symbolic calculation dealing with loop integrals (in
$n$-dimension), we extract the regulator constant
$C_{UV}=1/\varepsilon -\gamma_E+\log 4\pi$, $n=4-2 \varepsilon$
and treat this as a parameter. The ultraviolet finiteness test is
performed by varying the dimensional regularisation parameter
$C_{UV}$. This parameter could then be set to $0$ in further
computation. The test on the infrared finiteness is performed by
including both loop and bremsstrahlung contributions and checking
that there is no dependence on the fictitious photon mass
$\lambda$. An additional stability test concerns the
bremsstrahlung part. It relates to the independence in the
parameter $k_c$ which is a soft photon cut parameter that
separates soft photon radiation  and the hard photon performed  by
the Monte-Carlo integration. A crucial test concerns the gauge
parameter independence of the results. Gauge parameter
independence of the result is performed through a set of five
gauge fixing parameters. For the latter a generalised non-linear
gauge fixing condition\cite{nlg-generalised,nlgfatpaper} has been
chosen.

\beqn
\label{fullnonlineargauge} {{\cal L}}_{GF}&=&-\frac{1}{\xi_W}
|(\partial_\mu\;-\;i e \tilde{\alpha} A_\mu\;-\;ig c_W
\tilde{\beta} Z_\mu) W^{\mu +} + \xi_W \frac{g}{2}(v
+\tilde{\delta} H +i \tilde{\kappa} \chi_3)\chi^{+}|^{2} \nonumber \\
& &\;-\frac{1}{2 \xi_Z} (\partial.Z + \xi_Z \frac{g}{ 2 c_W}
(v+\tilde\varepsilon H) \chi_3)^2 \;-\frac{1}{2 \xi_A} (\partial.A
)^2 \;. \eeqn The $\chi$ represent the Goldstone. We take the 't
Hooft-Feynman gauge with $\xi_W=\xi_Z=\xi_A=1$ so that no
``longitudinal" term in the gauge propagators contributes. Not
only this makes the expressions much simpler and avoids
unnecessary large cancelations, but it also avoids the need for
high tensor structures in the loop integrals. The use of the five
parameters, $\tilde{\alpha}, \tilde{\beta}, \tilde{\delta},
\tilde{\kappa}, \tilde\varepsilon $ is not redundant as often
these parameters check complementary sets of diagrams.  Let us
also point out that when performing this check we keep the full
set of diagrams including couplings of the Goldstone and Higgs to
the electron for example, as will be done for the process under
consideration. Only at the stage of integrating over the phase
space do we switch these negligible contributions.

Although the system is not fully adapted for the computation of
generic QCD corrections, it is quite straightforward to implement
the QCD (final state) radiative corrections to \eettht. Indeed
these corrections are rather QED-like corrections. The infrared
divergence can be treated by giving the gluon an infinitesimal
mass while the ultraviolet divergences are treated via dimensional
regularisation. Also here we adopt an on-shell scheme in
particular for the top mass and wave function renormalisation. The
QCD $t\bar t H$ counterterm is then defined in terms of the top
mass counterterm and the wave function constant. We have checked
the infrared and ultraviolet finiteness of the QCD part also.

\begin{figure*}[htbp]
\begin{center}
\includegraphics[width=14cm,height=6.5cm]{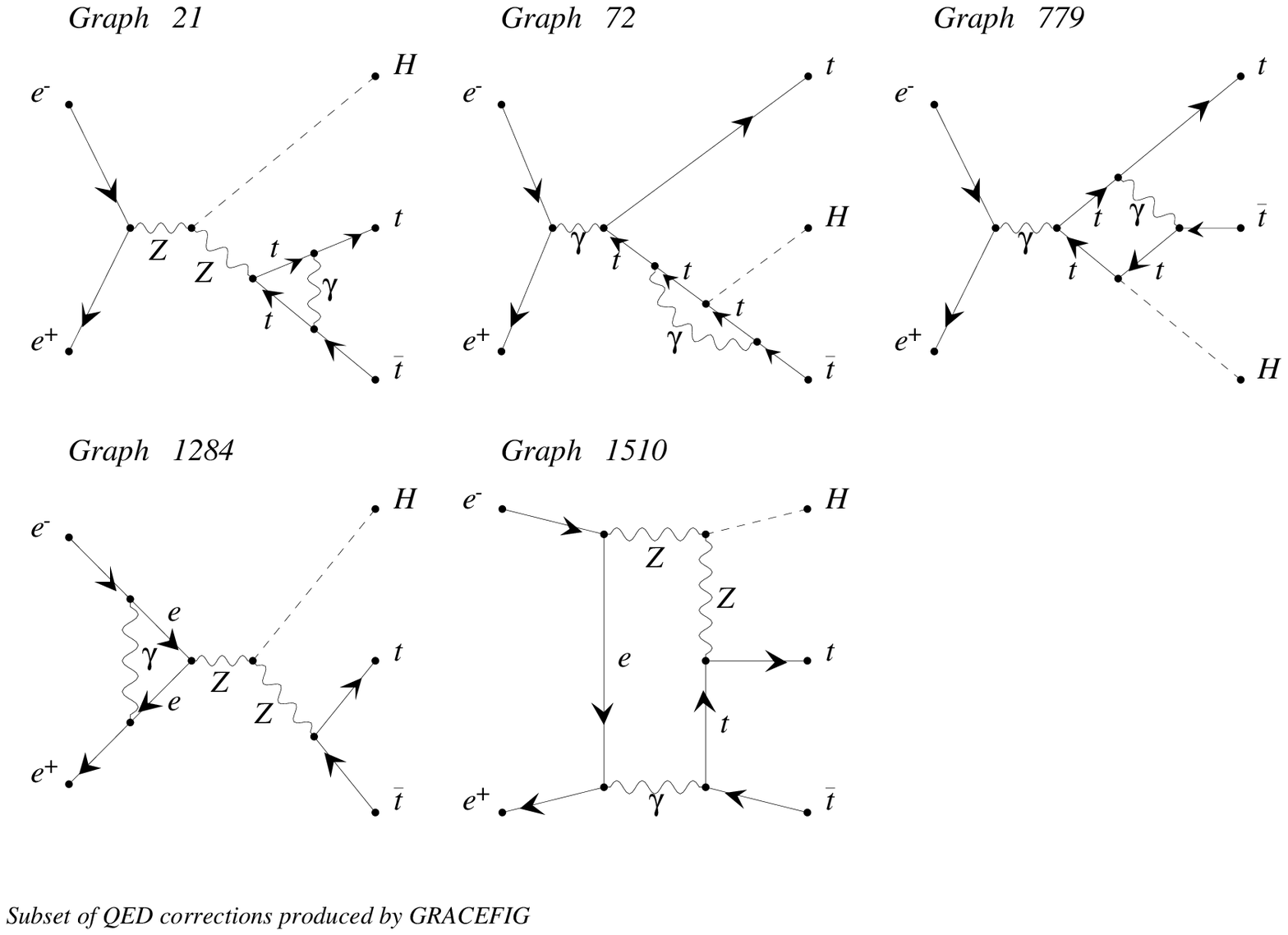}
\includegraphics[width=14cm,height=9.5cm]{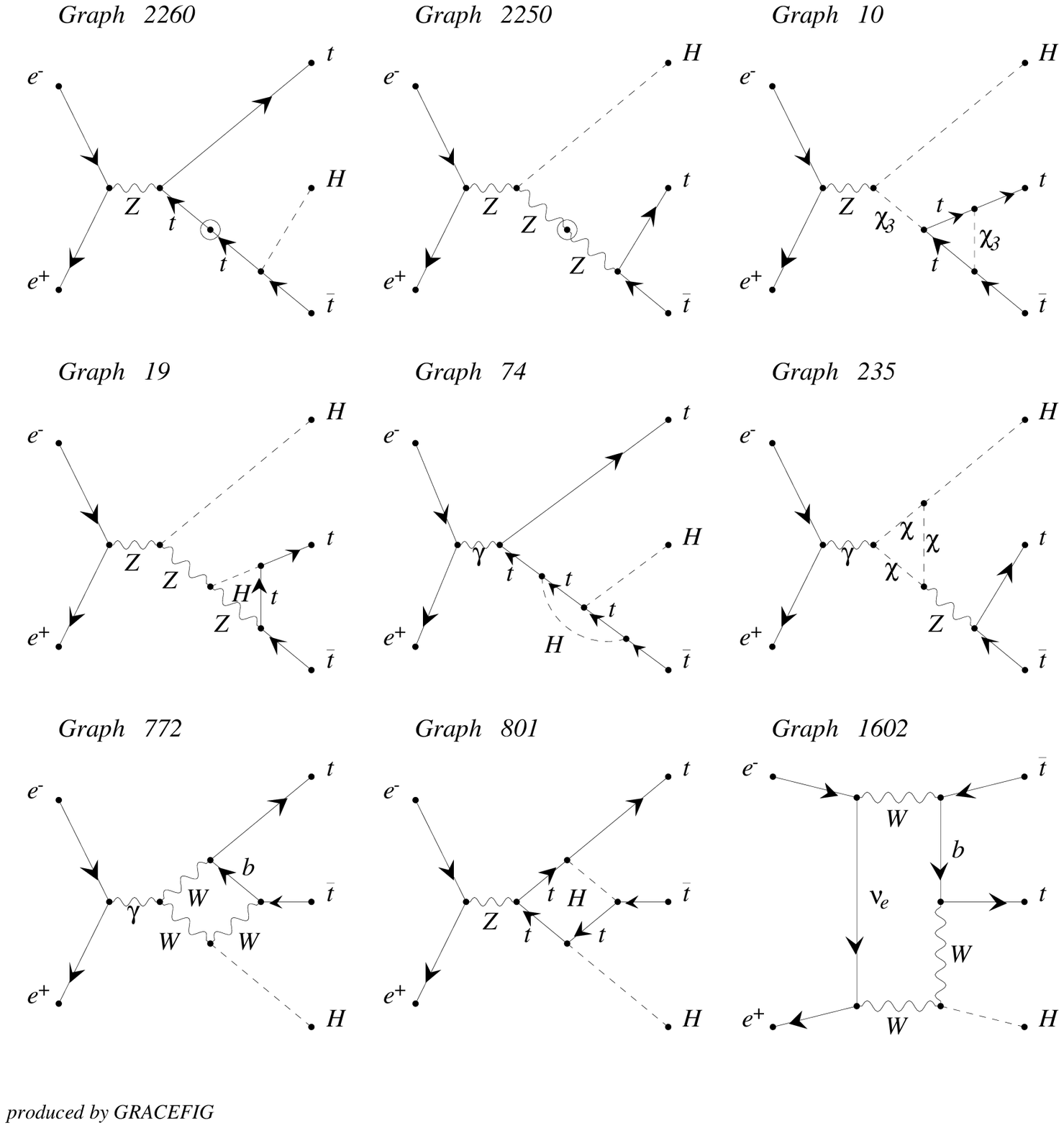}
\caption{\label{diagrams} {\small {\em A small selection of
different classes of loop diagrams contributing to \eetth. We keep
the same graph numbering as that produced by the system. The first
class of diagrams (the first 5 diagrams)  are QED corrections,
with the first row consisting of final state corrections. The QCD
corrections diagrams can be inferred from the latter. The pentagon
is a QED initial-final state interference. The second class groups
genuine electroweak corrections including self-energy, triangle,
box and pentagon corrections. Note that for the self-energy
diagrams we collect all contributions including the counterterms
in the blob. Note that the top self-energy {\tt Graph 2260}  also
contains a photonic correction that should be included in the
first class of diagrams.}}}
\end{center}
\end{figure*}

The full set of the Feynman diagrams within the non-linear gauge
fixing condition consists of 12 tree-level diagrams and 2327
one-loop diagrams (with 164 pentagon diagrams) for the electroweak
${\cal O}(\alpha)$ correction to the process $e^+e^-\rightarrow
t\bar{t}H$, see Fig.~\ref{diagrams} for a selection of these
diagrams. Even though we neglect the electron-Higgs coupling, the
set of diagrams still includes 6 tree-level diagrams and 758
one-loop diagrams (with 29 pentagon diagrams). We define this set
as the production set. To obtain the results of the total cross
sections, we use this production set. The handling of both the
scalar and tensor pentagon integrals is done exactly along the
lines in \cite{eennhletter} as was used for the calculation of
\nnht.


Our input parameters for the calculation of $e^+e^-\rightarrow
t\bar{t}H$ are the following. We will start by presenting the
results of the electroweak corrections in terms of the fine
structure constant in the Thomson limit with
$\alpha^{-1}=137.0359895$ and the $Z$ mass $M_Z=91.187$ GeV. The
on-shell renormalization program uses $M_W$ as an input. However,
the numerical value of $M_W$ is derived through $\Delta
r$\cite{Hiokideltar} with $G_\mu=1.16639\times 10^{-5}{\rm
GeV}^{-2}$\footnote{The routine we use to calculate $\Delta r$ has
been slightly modified from the one used in our previous paper on
\nnht\cite{eennhletter} to take into account the new theoretical
improvements. It reproduces quite nicely the approximate formula
in \cite{deltarhollik-approx}.}. Thus, $M_W$ changes as a function
of $M_H$. For the  lepton masses we take $m_e=0.510999$ MeV,
$m_\mu=105.658389$ MeV and $m_\tau=1.7771$ GeV. For the quark
masses beside the top mass $M_t=174$ GeV, we take the set
$M_u=M_d=63$ MeV, $M_s=92$ MeV, $M_c=1.5$ GeV and $M_b=4.7$ GeV.
With this we find, for example, that $M_W=80.3759$ GeV for
$M_H=120$ GeV and $M_W=80.3469$ GeV for $M_H=180$ GeV. For the QCD
coupling, we choose $\alpha_s(M_Z)=0.118$ as an input and evaluate
$\alpha_s(M_t)=0.10754$ with the next-to-next-to-leading order
renormalization group equation.


As well known, from the direct experimental search of the Higgs
boson at LEP2, the lower bound of the \sm Higgs boson mass is
114.4 GeV\cite{mhlimit-direct}. On the other hand, indirect study
of the electroweak precision measurement suggests that the upper
bound of the \sm Higgs mass is about 200
GeV\cite{mhlimit-03-2003}. In this paper, we therefore only
consider a relatively light \sm Higgs boson and take the two
illustrative values $M_H=120$GeV and $M_H=180$GeV.

Let us first present some quantitative consistency tests on our
results. For the electroweak part, the ultraviolet finiteness test
gives a result that is stable over $20$ digits when one varies the
dimensional regularisation parameter $C_{UV}$. As for the gauge
parameter independence checks, our results are stable over $26$
digits when varying any of the non-linear gauge fixing parameters.
For the QED infrared finiteness test we  also find results that
are stable over $20$ digits when varying the fictitious photon
mass $\lambda$. As for the $k_c$ stability test our results are
consistent within a Monte-Carlo statistical error of  $0.02\%$.

We also checked the tests of the stability on the ultraviolet and
the infrared finiteness for the QCD calculation. The ultraviolet
finiteness test gives a result that is stable over 30 digits. The
sum of loop and bremsstrahlung contributions is stable over 13
digits when varying $\lambda_g$, the infrared gluon mass
regulator. $k_c$ independence is consistent with a Monte-Carlo
statistical error of $0.2\%$. In addition, we reproduced the
previous results of ref.\cite{eetthdawson} when setting to zero
the $Z$ exchange diagrams and also exactly reproduced  the results
of Dittmaier {\it et al.}\cite{eetthzerwas} with the full ${\cal
O}(\alpha_s)$ calculation within the on-shell renormalization
scheme.

\section{Results}
\subsection{Full ${\cal O}(\alpha)$ and ${\cal O}(\alpha_s)$ corrections}

\begin{figure*}[htbp]
\begin{center}
\mbox{\includegraphics[width=8cm,height=8.5cm]{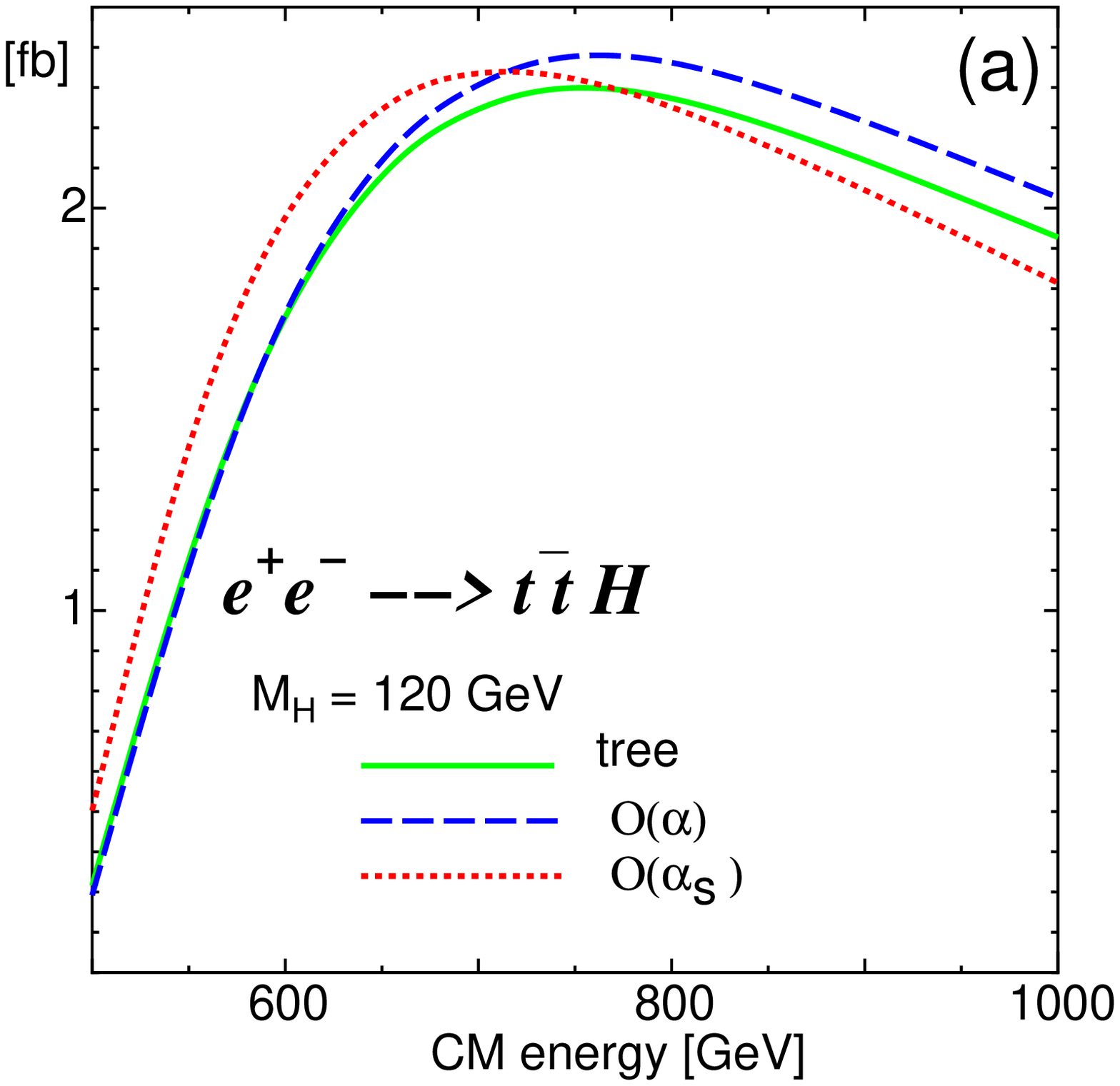}
\includegraphics[width=8cm,height=8.5cm]{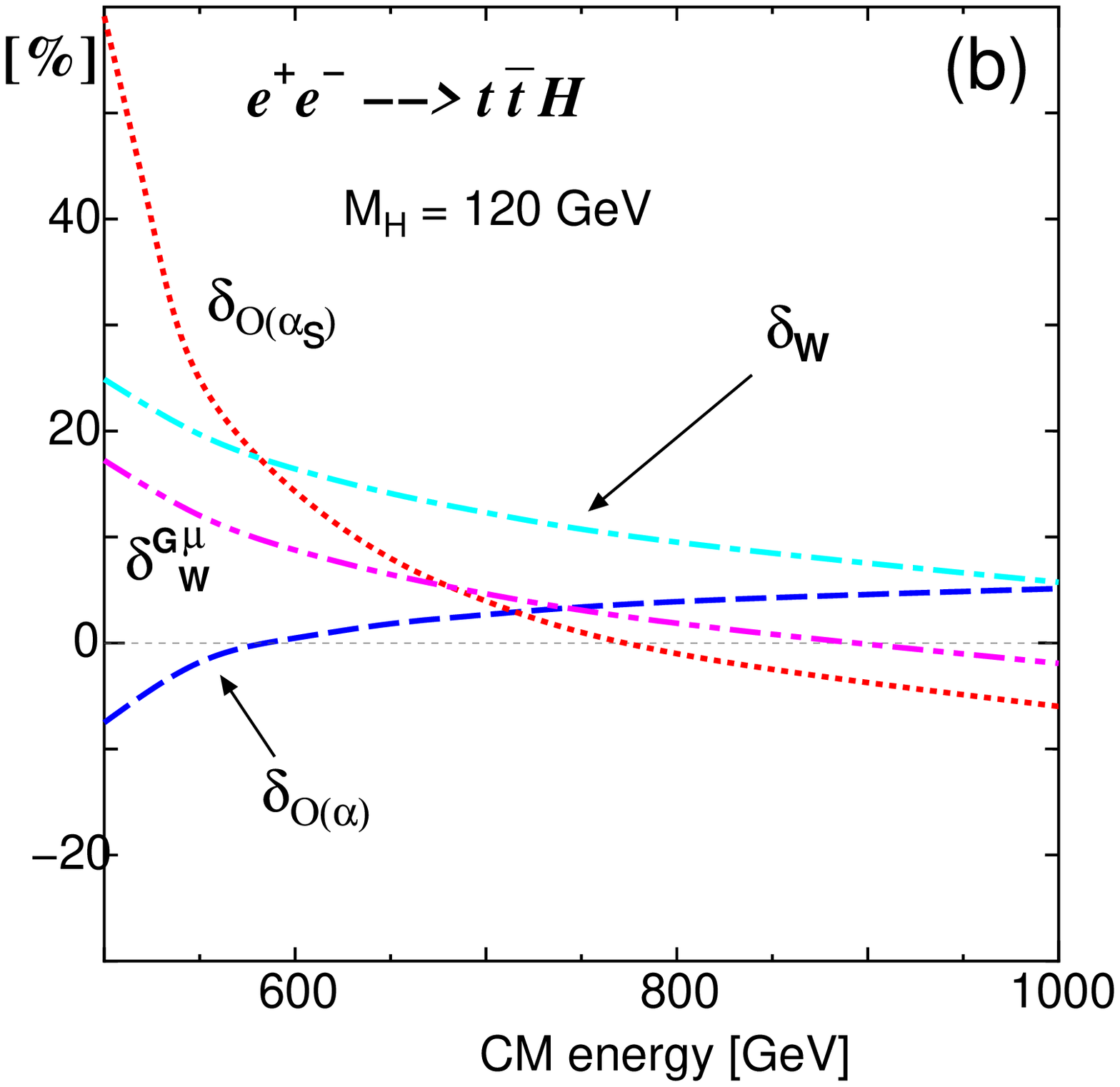}}
\caption{{\em (a) Total cross section as a function of the centre
of mass energy for $M_H=120$GeV. We show the total cross sections
for the tree level, full ${\cal O}(\alpha)$ and ${\cal
O}(\alpha_s)$ level in (a). The relative corrections are shown in
(b). Solid lines are tree level, dashed lines are the full ${\cal
O}(\alpha)$ and dotted lines are the  ${\cal O}(\alpha_s)$
corrections. We take $\alpha_s(M_t)=0.10754$. In addition, the
genuine weak correction $\delta_W$ and the relative correction
$\delta_W^{G_\mu}$ in the $G_\mu$ scheme are presented. }}
\label{CROSS.fig}
\end{center}
\end{figure*}

At tree-level, for the Higgs masses we are considering, the cross
section shows a steep rise just after threshold and slowly
decreases past the maximum of the cross section, see
Fig.~\ref{CROSS.fig}(a). For the measurement of the $t \bar t H$
coupling it is most useful to run at the maximum of the cross
section. For example for $M_H=120$GeV this maximum occurs around
$\sqrt{s}=700-800$GeV where the cross section is in excess of
$2fb$, see Fig.~\ref{CROSS.fig}(a). For a total integrated
luminosity of $1$ab$^{-1}$ the $1\sigma$ statistical error
corresponds to about a $2\%$ precision. Thus the theoretical
knowledge of the cross section at $0.2\%$ is quite sufficient. It
is important to keep in mind that the dominant contribution to
\eetth$\;$ for the energies we are considering is due to the
photon exchange diagram, the $Z$ exchange diagrams with Higgs
radiation off the $Z$ or the top is much smaller. This is
especially true at threshold. For example for $M_H=120$GeV and
$\sqrt{s}=500$GeV, the photon exchange diagram alone contributes
$90\%$ of the total cross section whereas Higgs radiation off the
$Z$ is less than $0.2\%$. The importance of the photon exchange
contribution lessens somehow at high energies, but even at $1$TeV
this contribution accounts for about $80\%$ of the total.

The full ${\cal O}(\alpha)$ electroweak correction and the full
${\cal O}(\alpha_s)$ QCD inclusive cross section take into account
the  full one-loop virtual corrections as well as the soft and
hard bremmstrahlung contributions. These corrections  are shown
Fig.~\ref{CROSS.fig}(a). The relative corrections defined as
\[
\delta_{{\cal O}(\alpha,\alpha_s)}= {\sigma_{{\cal
O}(\alpha,\alpha_s)} \over \sigma_{tree}}-1
\]
are shown in Figure \ref{CROSS.fig}(b).

Let us first turn to the inclusive QCD correction. It is useful to
write this correction as \beqn \label{CF} \delta_{{\cal
O}(\alpha_s)}= C_F \;\alpha_s(\mu) \; \Delta_s, \quad C_F=4/3.
\eeqn The QCD correction has a scale dependence which however, at
this order, is fully contained in $\alpha_s(\mu)$. We will provide
the exact value of $\alpha_s(\mu)$ together with $\delta_{{\cal
O}(\alpha_s)}$ so that a comparison with other calculations is
straightforward. As already noted in
\cite{eetthdawson,eetthzerwas} we confirm that the QCD corrections
are quite large at threshold increasing the tree-level cross
section by about $50\%$, for $M_H=120$GeV and $\sqrt{s}=500$GeV.
This large increase is due essentially to the gluonic Coulomb
corrections. At energies where the  cross section is at its
highest and the process is most likely to be of interest, the QCD
corrections are quite modest. For example, for $M_H=120$GeV and
$\sqrt{s}=800$GeV, as seen also in Table~\ref{QCDtab}, the
correction is a mere $\sim -1\%$, the residual scale dependence is
very small. Indeed as  shown in Table~\ref{QCDtab}, the
corrections for  a scale $\mu=M_t$ and $\mu=\sqrt{s}=800$GeV$\sim
4 M_t$, vary from $-1.$ to $-0.8\%$. A similar trend is observed
for $M_H=180$GeV, see Table~\ref{QCDtab}. As the energy increases
further past the maximum of the cross section, these corrections
turn negative but remain moderate at around $-5\%$ for
$\sqrt{s}=1$TeV. At these energies for the light Higgs masses we
are considering the $M_H$ dependence of the corrections lessens.
%

\begin{table*}[thb]
\begin{center}
\begin{tabular}{|c|c|c|lcc|}
\hline \rule[-1mm]{0mm}{5mm} $\sqrt{s}$ &
$M_H$(GeV)&$\sigma_{tree}$(fb)& $\sigma_{{\cal
O}(\alpha_s(M_t))}$(fb) & $\delta_{{\cal O}(\alpha_s(M_t))}(\%)$ &
$\delta_{{\cal
O}(\alpha_s(\sqrt{s}))}(\%)$ \\
\hline \rule[-1mm]{0mm}{5mm} 600 GeV & 120
          & $1.7293\pm0.0003$& $1.977\pm0.001$
              & $14.3$ & $12.4$~~
              \\
\rule[-1mm]{0mm}{5mm} ~       & 180
            & $0.33714\pm0.00004$      & $0.4383\pm0.0004$
              & $30.0$~~ & $26.0 $~~
\\
\hline \rule[-1mm]{0mm}{5mm} 800 GeV & 120
           &$2.2724\pm0.0005$ & $2.250\pm0.002$
              &  $-1.0$~~
              & $-0.8 $~~
\\
\rule[-1mm]{0mm}{5mm} ~       & 180
       & $1.0672\pm0.0003$  & $1.0856\pm0.0007$
        & $1.7$~~
        & $1.4$~~
\\
\hline \rule[-1mm]{0mm}{5mm} 1 TeV  & 120
     &$1.9273\pm0.0005$   & $1.812\pm0.003$
       & $-6.0$~~
     & $-4.9$~~
\\
\rule[-1mm]{0mm}{5mm} ~      & 180
      &$1.1040\pm0.0003$ & $1.049\pm0.001$
       & $-5.0$~~
       & $-4.1 $~~
\\
\hline
\end{tabular}
\caption{{\em QCD corrections for $e^+e^- \rightarrow t\bar{t}H$.
We also display the Monte-Carlo integration errors. We consider
two different schemes. (A) We choose the renormalization scale
$\mu$ of the QCD coupling $\alpha_s$ at $M_t=174$ GeV with
$\alpha_s(M_t)=0.10754$. (B) We take $\mu=\sqrt{s}$ with
$\alpha_s=0.09330$ at $\sqrt{s}=600$ GeV, $\alpha_s=0.09051$ at
$\sqrt{s}=800$ GeV and $\alpha_s=0.08847$ at $\sqrt{s}=1$TeV.}}
\label{QCDtab}
\end{center}
\end{table*}

We now turn to the total ${\cal O}(\alpha)$ electroweak
corrections. Again taking as an example $M_H=120$GeV, the ${\cal
O}(\alpha)$ electroweak correction is about $-7.5$\% around
threshold at $\sqrt{s}=500$ GeV and is therefore swamped by the
QCD correction. However as the energy increases, contrary to the
QCD correction, the full electroweak correction  slowly increases
and turns positive for $\sqrt{s}>600$ GeV. Around the maximum of
the cross section at $\sqrt{s}=800$, the full electroweak
correction is about $+4\%$ and thus larger than the full QCD
correction ($-1\%$). For yet higher energies these corrections
tend to cancel each other. For example for $M_H=120$GeV this
occurs around $\sqrt{s}=1$TeV. We also show some numerical results
$M_H=180$ GeV in Table~\ref{EWtab}. The ${\cal O}(\alpha)$
corrections for the Higgs mass of 180GeV are rather small in the
range $\sqrt{s}=800$ GeV $-~1$ TeV, slightly increasing from $\sim
-2\%$ to $-0.5\%$. In particular,  for $\sqrt{s}=1$TeV and
$M_H=180$GeV the full ${\cal O}(\alpha)$, because of its
smallness, may not be resolved from the corresponding QCD
corrections due to the scale dependence of the latter.

\begin{table*}[thb]
\begin{center}
\begin{tabular}{|c|c|ccr|}
\hline \rule[-1mm]{0mm}{5mm} $\sqrt{s}$ & $M_H$(GeV)&
$\sigma_{tree}$(fb) & $\sigma_{{\cal O}(\alpha)}$(fb)&
$\delta_{{\cal O}(\alpha)}(\%)$ \\
\hline \rule[-1mm]{0mm}{5mm} 600 GeV & 120 & $1.7293\pm0.0003$  &
$1.738\pm0.002$
              & $0.5$~~\\
\rule[-1mm]{0mm}{5mm} ~       & 180 & $0.33714\pm0.00004$ &
$0.3126\pm0.0003$
              & $-7.3$~~\\
\hline \rule[-1mm]{0mm}{5mm} 800 GeV & 120 & $2.2724\pm0.0005$ &
$2.362\pm0.004$
              & $3.9$~~\\

\rule[-1mm]{0mm}{5mm} ~       & 180 & $1.0672\pm0.0003$ &
$1.050\pm0.002$
        & $-1.6$~~\\
\hline \rule[-1mm]{0mm}{5mm} 1 TeV  & 120 & $1.9273\pm0.0005$ &
$2.027\pm0.004$
       &  $5.2$~~\\
\rule[-1mm]{0mm}{5mm} ~      & 180 & $1.1040\pm0.0003$ &
$1.098\pm0.002$
       & $-0.5$~~\\
\hline
\end{tabular}
\caption{{\em  As in the previous table but for the total ${\cal
O}(\alpha)$ electroweak corrections. }} \label{EWtab}
\end{center}
\end{table*}

\subsection{The genuine weak correction}
In order to quantify the genuine weak corrections one needs to
subtract the full QED corrections from the full ${\cal O}(\alpha)$
corrections. This is important because  it is well known that the
QED corrections can be quite large and that in \epemt processes
those from the initial state need to be resummed\cite{qedps}. For
the process at hand,  which at tree-level proceeds through
$s$-channel neutral vector bosons, these QED corrections form a
gauge invariant set. This set may be further subdivided into three
subsets that are also separately gauge invariant: {\it i)} initial
state radiation, {\it ii)} purely final state radiation and {\it
iii)} the initial-final state QED interference.

\noi {\it i)} The dominant initial state QED virtual and soft
bremsstrahlung corrections are given by the universal soft photon
factor that leads to a relative correction\cite{eennhletter} \bea
\label{dqeduniv} \delta_{V+S,in.}^{QED}=\frac{2
\alpha}{\pi}\left((L_e-1)\ln \frac{k_c}{E_b}+\frac{3}{4}L_e +
\frac{\pi^2}{6}-1 \right), \quad \; L_e=\ln(s/m_e^2) \;. \eea \noi
where $m_e$ is the electron mass, $E_b$ the beam energy
($s=4E_b^2$) and $k_c$ is the cut on the soft photon energy.

\noi {\it ii)} The total QED final state radiation can also be
read off from the result of the QCD radiative correction through
the replacement $\alpha_s(\mu)\; C_F  \ra \alpha \; Q_t^2 $ in
Eq.~\ref{CF} ($Q_t$ is the electric charge of the top).

\noi {\it iii)} The initial-final state QED correction is
ultraviolet finite. Within our system this contribution can be
easily isolated and combined with the appropriate bremmstrahlung
counterpart leading to an infrared finite result.

Although this approach of extracting the full QED correction is
the most simple one, we have also calculated the full QED
corrections separately  and subtracted their contributions from
the full ${\cal O}(\alpha)$. In order to perform this subtraction,
the QED virtual corrections are generated by dressing the
tree-level diagrams with one-loop photons (the photon self-energy
is not included in this class). Moreover one needs to include some
counterterms. One only has to take into account the purely
photonic contribution to the top mass counterterm  as well as  the
wave function renormalisation constants of the electron and the
top. Performing this more direct computation, we confirmed, that
especially around threshold, to a large extent the bulk of the QED
corrections originate from the initial state universal
corrections. Moreover this also checked the extraction of the
final QED corrections. A break up of the soft and virtual QED
corrections into initial, final and interference is shown in
Table~\ref{qedtab}.

\begin{table*}[hbtp]
\begin{center}
\begin{tabular}{|c||c|c|c|c|c||c|c|}
\hline
$M_H$&$\sqrt{s}$&$\sigma^{QED}_{V+S,Full}$(fb) &$\sigma^{QED}_{V+S,Init.}$&$\sigma^{QED}_{V+S,Fin.}$ &$\sigma^{QED}_{V+S,Int.}$&$\delta^{QED} $& $\delta_W$\\
(GeV)&(GeV) &($\sigma^{QED}_{hard,Full}$(fb))&(fb)&(fb)&(fb)&$(\%)$& $(\%)$\\
\hline
$120$&&&& &&&\\
   &600 & -2.6092  & -2.557
   &-0.012&-0.043&-16.0&16.5\\
   & & (2.3333)&&&&&\\
   &800  &-3.6667 &  -3.516
   &-0.055&-0.099&-5.6&9.5\\
      & & (3.5391)&&&&&\\
  &1000  &-3.2622&  -3.086
  & -0.071&-0.109&-0.6&5.8\\
     & & ( 3.2507)&&&&&\\
 \hline
$180$&&&&&&&\\
   &600  &-0.50490& -0.4985
   & 0.0056&-0.0072&-25.7&18.4\\
      & & ( 0.41839)&&&&&\\
   &800 & -1.7120& -1.651
   &-0.020&-0.043&-10.7&9.1\\
      & & ( 1.5975)&&&&&\\
   &1000  &-1.8589 &  -1.768
   &-0.035&-0.058&-4.9&4.4\\
      & & (1.8048)&&&&&\\
   \hline
\end{tabular}
\caption{{\em  Extraction of the QED corrections.
$\sigma^{QED}_{V+S,Full}$ corresponds to the cross section for the
full one-loop QED virtual and soft bremsstrahlung with
$k_c=0.001$GeV. $\sigma^{QED}_{V+S,Init.}$ extracts the initial
state radiation. $\sigma^{QED}_{V+S,Fin.}$ gives the final state
QED correction whereas $\sigma^{QED}_{V+S,Int.}$ is the
initial-final QED interference contribution. We also give
$\sigma^{QED}_{hard,Full}$ which is the full hard photon radiation
cross section. All cross sections are in fb. We also give the
relative  QED correction (after including hard radiation) as well
as the relative genuine weak correction as defined in the text.
Note also that the extraction of the total
$\sigma^{QED}_{V+S,Full}$ and $\sigma^{QED}_{hard,Full}$ has been
computed with higher accuracy than the  individual (S+V)
contributions. We can check that the two computations (full) and
adding the individual contributions agree.}} \label{qedtab}
\end{center}
\end{table*}
\noi We define the genuine weak relative correction as,
\[
 \delta_W=\delta_{{\cal O}(\alpha)}-\delta^{QED}= \delta_{{\cal O}(\alpha)}-\delta_{V+S}^{QED}
- \delta_{hard}^{QED}.
\]
$\delta_{V+S}^{QED}$ is the complete QED virtual and soft
correction whereas $\delta_{hard}^{QED}$ is the hard photon
contribution. The weak corrections after subtraction of the QED
corrections are shown in Fig.~\ref{CROSS.fig} for $M_H=120$GeV.
The total genuine weak corrections are not small, being largest
around threshold ($\sim +25\%$ at $\sqrt{s}=500$GeV) and decrease
monotonically as the energy increases, see
Fig.~\ref{CROSS.fig}(b). At $\sqrt{s}=1$TeV they reach about
$6\%$. These corrections could therefore, for this Higgs mass,
always be disentangled from the QCD corrections. Past
$\sqrt{s}=600$GeV and up to $\sqrt{s}=1$TeV , they are larger, in
absolute terms, than the QCD corrections. A similar trend also
occurs for $M_H=180$GeV, see Table~\ref{qedtab}. For
$\sqrt{s}=600$GeV we find $\delta_W \sim +18\%$ this correction
drops with energy reaching $\delta_W \sim  +4\%$ at $1$TeV where
it almost cancels the
corresponding QCD correction. \\

Having subtracted the genuine weak corrections one could also
express the corrections in the $G_\mu$ scheme by further
extracting the rather large universal weak corrections that affect
two-point functions through $\Delta r$. This defines the genuine
weak corrections in the $G_\mu$ scheme as
$\delta_W^{G_\mu}=\delta_W-3\Delta r$. For reference, one has
$\Delta r=2.55\%$ for $M_H=120$ GeV and $\Delta r=2.70\%$ for
$M_H=180$ GeV. For \nnht this procedure helps absorb a large part
of the weak corrections. Another advantage is that much of the
(large) dependence due to the light fermions masses also drops
out. For $M_H=120$GeV, the relative correction $\delta_W^{G_\mu}$
is shown as a function of energy in Fig.~\ref{CROSS.fig}. Adopting
this scheme, we find that in the energy range where the cross
section is largest, $\sqrt{s}=700$GeV to $\sqrt{s}=1$TeV, the
correction remains modest changing from about $5\%$ to $-2\%$.
These corrections could be ``measured above" the QCD corrections.
For $M_H=180$GeV, the correction ranges from $1\%$ to $-4\%$ as
the energy changes from $800$GeV to $1$TeV.  However at energies
around threshold, the genuine weak corrections in the $G_\mu$
scheme  are large (and positive), although about a factor $3$ to
$2$ smaller than the QCD corrections. For $M_H=120$GeV the genuine
non-QED correction in the $G_\mu$ scheme reaches $+17\%$ at
$\sqrt{s}=500$GeV ($\sim 30$GeV above threshold). They are about
$10\%$ for $M_H=180$GeV $\sqrt{s}=600$GeV ($\sim 70$GeV above
threshold). These corrections slightly decrease with energy
following a trend similar to the QCD corrections although the
decrease is not as fast. One knows from $\epem \ra t \bar t$
\cite{tt-threshold,eettrcfujimoto} that there is a Yukawa
counterpart to the Coulomb large gluonic correction mediated by a
Higgs exchange which is large at threshold for a large top mass
and a very light Higgs mass. This phenomenon can only partially
account for the large $17\%$ increase of the cross section at
$\sqrt{s}=500$GeV for $M_H=120$GeV, considering the value of the
Higgs mass. We would like to argue that because the tree-level
cross section is dominated by photon exchange, a description of
the cross section in terms of $G_\mu$ instead of $\alpha$ may not
be the most appropriate. Rather, the photon couplings  should be
parameterized  in terms of the running $\alpha$. Therefore
especially at threshold one should subtract from $\delta_W$
$2\Delta \alpha(s)+\Delta r$ instead of $3\Delta r$. For
illustration we will only use the light fermion contribution to
$\Delta \alpha(s)$. The correction defined this way,
$\delta_{mixed}=\delta_W-2\Delta \alpha(s)-\Delta r$, brings down
the remaining weak contribution to $\delta_{mixed}\sim 7\%$ for
$M_H=120GeV$ at $\sqrt{s}=500$GeV and $1.4\%$ for $M_H=180$GeV at
$\sqrt{s}=600$GeV. For higher energies, above $\sqrt{s}=700$GeV,
this prescription gives large (negative) corrections. For
instance, for $\sqrt{s}=1$TeV we find $\delta_{mixed}\sim -12\%$
for $M_H=120GeV$ and $\delta_{mixed}\sim -14\%$ for $M_H=180GeV$.
Therefore none of the prescriptions, $G_\mu$ or $\delta_{mixed}$
reproduces the full genuine weak corrections across the whole
energy range from threshold to $1$TeV. This should not be
surprising as it has been known, already for $2\ra 2$ processes,
that the contribution of boxes becomes important as the energy
increases and that the $G_\mu$ scheme is not always the most
appropriate. A case in point is \eezht\cite{eezhsmrc}. One may
also enquire whether  the leading $m_t^2$ corrections to the $t
\bar t H$ coupling could account for most of the weak radiative
corrections. The leading $m_t^2$ corrections to the $f \bar f H$
vertex, in the $G_\mu$ scheme, had  been worked out
\cite{HiokiffH,WWH-Yukawa}. The corrected $t\bar t H$ vertex
$y_{t\bar t H}$ writes in terms of the tree-level one, $y_{t\bar t
H}^0$, as \beqn y_{t\bar t H}=y_{t\bar t H}^0 \left(1+\frac{7}{2}
\frac{G_\mu M_t^2}{8 \pi^2 \sqrt{2}}\right) \eeqn

\noi In the total cross section with $M_t=174$GeV this only
accounts for about $2.2\%$ weak correction. Therefore one sees
that to properly take into account the electroweak corrections to
the \eetth$\;$ a full calculation is needed.

\section{Conclusions}

We have performed a full one-loop correction to the process
$e^+e^- \rightarrow t\bar{t}H$ which, at a future linear collider
running in the energy range $700$GeV to $1$TeV, can allow a direct
determination of the important Yukawa coupling $t\bar t H$. The
full one-loop corrections combine both the full electroweak
corrections as well as the QCD corrections. Apart from the
ultra-violet and infrared finiteness tests of the results, we
performed, for the more involved electroweak sector, an extensive
gauge parameter independence check. $k_c$ stability has also been
verified. $k_c$ is the photon energy parameter that separates soft
and hard photon (and gluon) radiation. For the electroweak part we
have also extracted the contribution of the QED corrections. The
final state QED correction has also been used as a further check
on the QCD part. This extraction helps define the genuine weak
corrections. We have expressed the latter both in the
$\alpha$-scheme and the $G_\mu$-scheme. We find that, for all the
Higgs masses that we have studied, the full ${\cal O}(\alpha)$
corrections are swamped by the large QCD corrections at threshold.
The latter can increase the cross section by as much as $50\%$ due
to the threshold Coulomb enhancement. It is interesting to note
that after the extraction of the QED corrections, the genuine weak
corrections at threshold are important ($\sim 20\%$), although
still small compared to the corresponding QCD corrections.
However,  around the peak of the cross section where this process
is most likely to be of interest for the extraction of the Yukawa
coupling, the electroweak corrections can dominate over the QCD
corrections. In this energy range both corrections are under
control and for $M_H=120$GeV, the genuine weak corrections
expressed in the $G_\mu$ scheme are modest ranging between $5\%$
to $-2\%$ for $\sqrt{s}=500$GeV to $\sqrt{s}=1$TeV. Nonetheless
they have to be taken into account for a precision analysis of
this process at a future linear collider.

\vspace{1cm} {\bf Acknowledgments} \\
\noi This work is part of a collaboration between the {\tt GRACE}
project in the Minami-Tateya group and LAPTH. D. Perret-Gallix and
Y. Kurihara deserve special thanks for their contribution. We also
acknowledge useful discussions with J.P.~Guillet and E.~Pilon.
This work was supported in part by the Japan Society for Promotion
of Science under the Grant-in-Aid for scientific Research
B(N$^{{\rm o}}$. 14340081) and PICS 397 of the French National
Centre for Scientific Research (CNRS).

\vspace{1cm} {\bf Addendum}\\
\noi Our results for the full ${\cal O}(\alpha)$ corrections were
first reported in \cite{eetthloopfest2}. In the meantime while
finalising this article there appeared a paper on the same
subject\cite{eetthchinese}. These authors calculate the ${\cal
O}(\alpha)$ corrections using the system {\tt FeynArts} and {\tt
FeynCalc}\cite{FeynArts} but do not attempt to isolate the genuine
weak corrections. These authors use a rather different set of
input parameters. For the light quark masses they use the current
quark masses which would give a much too large value of $\Delta
\alpha (M_Z^2)$. Nonetheless taking the same input parameters we
have run our program for a few points  reported
in\cite{eetthchinese}. With $M_H=115$GeV, for both
$\sqrt{s}=500$GeV and $\sqrt{s}=1000$GeV we find very satisfactory
agreement, at least as far as one can read from their graph
(Fig.~5 of \cite{eetthchinese}). We obtain
$-4.2\%$($\sqrt{s}=500$GeV) and $7.3\%$($\sqrt{s}=1$TeV). However
for higher energies we have not been able to reproduce their
results. For instance for $\sqrt{s}=2$TeV we find $\delta_{ {\cal
O}(\alpha)}\sim 8\%$ whereas they find $\sim 4\%$.\\

{\bf Note added}\\
\noi After submitting this paper there appeared another
calculation of this process\cite{Dennereetth}. This new
calculation perfectly confirms all of our results, for instance
the relative corrections in the two codes agree within $0.1\%$ at
all energies and for all masses.

\end{document}